%% file: QCD-10-023_temp.tex
\pdfoutput=1
\documentclass[11pt,twoside,a4paper,cmspaper,final,collab]{cms-tdr}
\def\svnVersion{32084M}\def\cmsCernNo{2010-091}\def\cmsCernDate{\today}\def\cmsMessage{Submitted to the Journal of High Energy Physics}

\begin{document}\cmsNoteHeader{QCD-10-023}
\hyphenation{env-iron-men-tal}%    just an example
\hyphenation{had-ron-i-za-tion}
\hyphenation{cal-or-i-me-ter}
\hyphenation{de-vices}
\RCS$Revision: 31986 $
\RCS$HeadURL: svn+ssh://alverson@svn.cern.ch/reps/tdr2/papers/QCD-10-023/trunk/QCD-10-023.tex $
\RCS$Id: QCD-10-023.tex 31986 2011-01-17 18:09:19Z paoloc $
\input{ptdr-definitions}
\providecommand{\dedx}{\ensuremath{\text{d}E/\text{d}x}\xspace}
\cmsNoteHeader{QCD-10-023} % This is over-written in the CMS environment: useful as preprint no. for export versions
\title{Measurement of Bose--Einstein Correlations in pp  Collisions at $\sqrt{s}=0.9$ and $7\TeV$ }
\address[pd]{INFN and University of Padova, Italy}
\address[cyp]{University of Cyprus, Cyprus}
\address[pau]{Istituto de Fisica Teorica, Universitade Estadual Paulista, Sao Paulo, Brazil}
\author[pd]{P.~Checchia}
\author[pd]{T.~Dorigo}
\author[cyp]{M.~Galanti}
\author[pd]{S.~Lacaprara}
\author[pd]{M.~Margoni}
\author[pd]{M.F.~Moro}
\author[pd]{M.~Nespolo}
\author[pau]{S.~Padula}
\author[pd]{L.~Perrozzi }
\author[pd]{F.~Simonetto}
\author[pd]{E.~Torassa}
\date{\today}
\abstract{
    Bose--Einstein correlations between identical particles are measured in samples
of proton-proton collisions at 0.9 and 7  TeV centre-of-mass energies, recorded
  by the CMS experiment at the
  LHC.
    The signal is observed in the form of an enhancement of number of pairs of
    same-sign charged particles with small relative momentum.
The dependence of this enhancement on kinematic and topological features of the event is studied.}
\hypersetup{%
pdfauthor={CMS Collaboration},%
pdftitle={Measurement of Bose-Einstein Correlations in pp  Collisions at sqrt(s)=0.9 and 7 TeV},%
pdfsubject={CMS Bose--Einstein},%
pdfkeywords={CMS, physics, QCD, low-pt, Bose--Einstein}}

\maketitle %maketitle comes after all the front information has been supplied
\section{Introduction \label{s:bec}}

In particle collisions, the space-time structure of the hadron emission region
can be studied using measurements of Bose--Einstein correlations (BEC) between pairs of identical bosons.
Since the first observation of BEC in proton-antiproton
interactions fifty years ago~\cite{goldhaber}, a large number of measurements have been performed by
experiments using different
initial states~\cite{alex,revbec}.
At the CERN Large Hadron Collider (LHC), BEC   were observed for the first time by CMS
using data at centre-of-mass energies $\sqrt{s} = 0.9$  and 2.36 TeV, collected  in 2009~\cite{bec_prl};
 measurements by ALICE at 0.9 TeV were  reported in \cite{ALICE_BEC}.
The present paper reports measurements using data taken  in 2010 at 0.9 TeV,
with a sample increase by a factor 15, and at 7 TeV, for the first time.
The analysis method is similar to that in~\cite{bec_prl}, where more details can be found.
In this article the results at the two energies  are compared and additional studies are performed.

Constructive interference affects the joint probability for the emission of a
pair of identical bosons with four-momenta $p_1$ and $p_2$.
Experimentally, the proximity in phase space between final-state particles
is quantified by
 the Lorentz-invariant quantity
$Q =\sqrt{-(p_1-p_2)^2}=\sqrt{M^2-4m_\pi^2}$, where $M$ is the invariant mass of the two particles,
assumed to be pions with mass  $m_{\pi}$.
The BEC effect is observed
 as an enhancement at low $Q$ of the ratio of the $Q$ distributions
for pairs of identical particles in the same event, to that
for pairs of particles in a reference sample that by construction is expected to include no BEC effect:
\begin{linenomath}
\begin{equation}
\label{eq:R1}
 R(Q) =  (\text{d}N / \text{d}Q)/ (\text{d}N_{\text{ref}} / \text{d}Q).
\end{equation}
The ratio is  fitted with the parameterization
\begin{equation}
\label{eq:exp}
 R(Q) = C\left[1+ \lambda \Omega(Qr)\right] \cdot (1+ \delta Q).
\end{equation}
\end{linenomath}
\noindent
In a static model of particle sources, $\Omega(Qr)$ is the Fourier transform
of the space-time region emitting bosons with overlapping wave
functions, characterized by an effective size $r$.
The parameter $\lambda$ measures the strength of BEC for incoherent boson
emission from independent sources,
$\delta$ accounts for long-distance correlations, and $C$ is a normalization factor.
The correlation function  is often  parameterized as an exponential $\Omega(Qr)=e^{-Qr}$
or with a Gaussian form $\Omega(Qr)=e^{-(Qr)^2}$. Other forms have also been used
(\cite{kozlov0} and references therein),
and several of these are mentioned below.  In addition a formulation
aimed at describing the time evolution of the source \cite{alevy}
is considered and compared with the data.
\section{Data and Track  Selection \label{s:datasets} }
 A detailed description
of the CMS detector can be found in~\cite{cmstdr}.
The central feature
of the CMS apparatus is a superconducting solenoid of 6 m internal diameter, providing an
axial magnetic field of 3.8 T.
The inner tracking system is the most relevant detector for the present analysis. It is composed
of a pixel detector with three barrel layers at radii between 4.4 and 10.2\cm and a silicon
strip tracker with 10 barrel detection layers extending outwards to a radius of 1.1~m.
Each system is completed by two endcaps, extending the acceptance up to a pseudorapidity  $|\eta|= 2.5$.
The transverse-momentum ($\pt$) resolution, for 1\GeV charged particles, is between $0.7\%$ at $\eta=0$ and $2\%$ at $|\eta|=2.5$.

Minimum-bias events were selected by requiring activity in both beam scintillator
counters~\cite{dNdEta}.
Charged particles are required to have $|\eta|<2.4$ and $\pt>200$\MeV,
ensuring that particles emitted from the interaction region cross all
three barrel layers of the pixel detector and thus have good two-track
separation.
To achieve a high purity of the primary track selection, the trajectories
are required to be reconstructed in fits
with more than five degrees of freedom ($N_{{\rm dof}}$) and $\chi^2/N_{{\rm dof}}<5.0$.
The transverse impact parameter with respect to the collision point is required
to be less than $0.15\cm$.
 The innermost measured point of the track must
be within $20\cm$ of the beam axis, in order to
reduce contamination from electrons and
positrons from photon conversions in the detector material and
secondary particles from  decay of long-lived hadrons.

For this analysis a total of 4.2 million events were selected at $\sqrt{s}=0.9$ TeV,
with 51.5 million tracks passing the selection criteria.
At 7 TeV, 2.7 million events with  51.7 million tracks were selected from data taken during
low-intensity runs.
Neither of the two energy samples is affected by event pileup.
Several minimum-bias Monte Carlo (MC) samples were
generated, followed by  detailed detector simulation based on the {\sc GEANT4} package~\cite{GEANT4}.
At 0.9 TeV, the MC simulations were generated with several {\sc PYTHIA6.4}~\cite{pythia64}  tunes
(D6T, DW, Perugia0, Z1 and Z2~\cite{MCtunes,Pro,P0}). At  7 TeV,
the simulations use {\sc PYTHIA6.4} tunes
(ProPt0, Perugia0, Z1 and Z2) and  {\sc PYTHIA8.1}~\cite{pythia8}.

\section{ Definition of Signal and Reference Samples}
All pairs of same-sign charged particles with
$Q$ between $0.02$ and $2\GeV$
are used for the measurement.
The lower limit is chosen to avoid cases of tracks that are duplicated or not well separated, while
the upper limit extends far enough beyond the signal region (confined to $Q<0.4$ GeV)
to allow verification of a good match between signal
and reference samples. The $Q$ resolution in the signal region  is
better than 10 MeV.
Coulomb interactions between charged particles modify their relative momentum distribution.
This effect, which differs for pairs with same charge (repulsion) and opposite charge (attraction),
is corrected
using Gamow factors~\cite{noGamov}.
As discussed in~\cite{bec_prl}, the reference sample in the denominator
of Eq.~(\ref{eq:R1}) can be defined in several ways:
{\it opposite-charge pairs};
{\it  opposite-hemisphere pairs}, where  particles are
paired after inverting  the three-momentum of one of them,
this procedure
being applied to pairs with same and opposite
charges;
{\it rotated particles}, where  pairs are constructed by inverting
the $x$ and $y$ components
of the three-momentum of one of the two particles;
{\it pairs from mixed events}.
In the case of pairs from mixed events, particles from different events are combined with the
following methods: i) events are mixed at random; ii) events with similar
  charged-particle multiplicities in the same $\eta$ regions are selected;
  iii) events with an invariant mass of all charged particles similar
  to that of the signal are used to form the pairs.
In this paper, we use the sample obtained by pairing  same-sign charged particles from
different events that have similar charged-particle multiplicities in the same $\eta$ regions.
This method avoids the possible effect of remaining  correlations \cite{corr} between particles within the same event.
The r.m.s.~spread of the results obtained from the different samples
is taken as a conservative systematic uncertainty.
In \cite{bec_prl} an  additional, ``combined'' reference sample was  obtained by summing the $Q$ distributions of
the seven corresponding reference samples.
It has been checked that the
results obtained with the ``combined'' reference
sample are compatible within errors with those
presented here.

In order to reduce possible biases
in the construction of the reference sample, a double ratio
$\cal{R}$
 is defined,
\begin{linenomath}
\begin{equation}
\label{eq:doubleratio}
{\cal{R}}(Q) = \frac{R}{R_{\text{MC}}} =
\left(\frac {\text{d}N/\text{d}Q} {\text{d}N_{\text{ref}}/\text{d}Q} \right)
\Big/
\left(\frac {\text{d}N_{\text{MC}}/\text{d}Q}{\text{d}N_{\text{MC,\,ref}}/\text{d}Q}\right),
\end{equation}
\end{linenomath}
\noindent
where the subscripts ``MC''  and ``MC,\,ref'' refer to the corresponding distributions from the simulated events,
generated without BEC effects.
\section{Determination of Bose--Einstein Correlation Parameters \label{s:results}}

\begin{figure}[!h]
    \begin{center}
\includegraphics[width=0.35\textwidth, angle=90]{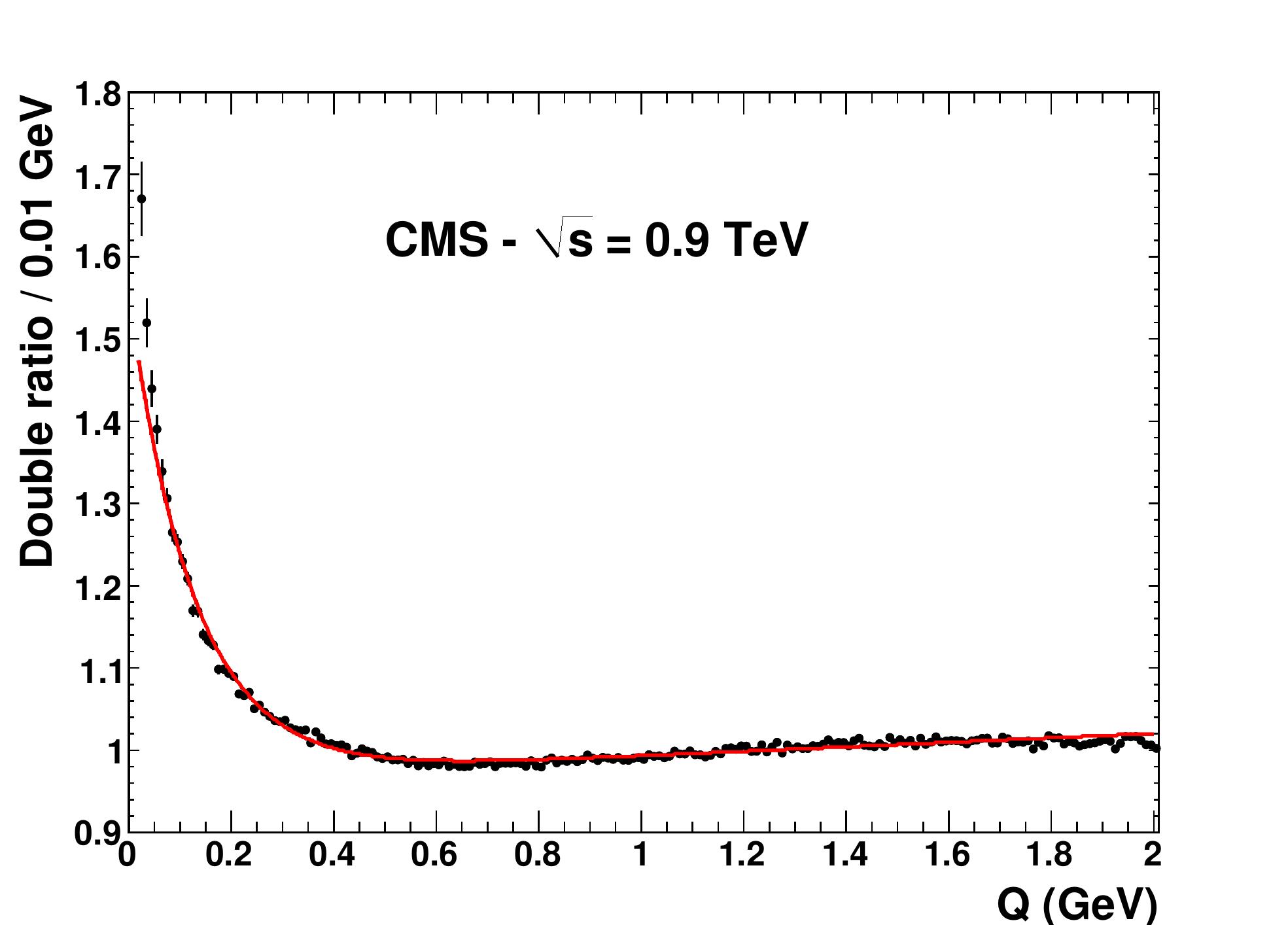}
\includegraphics[width=0.35\textwidth, angle=90]{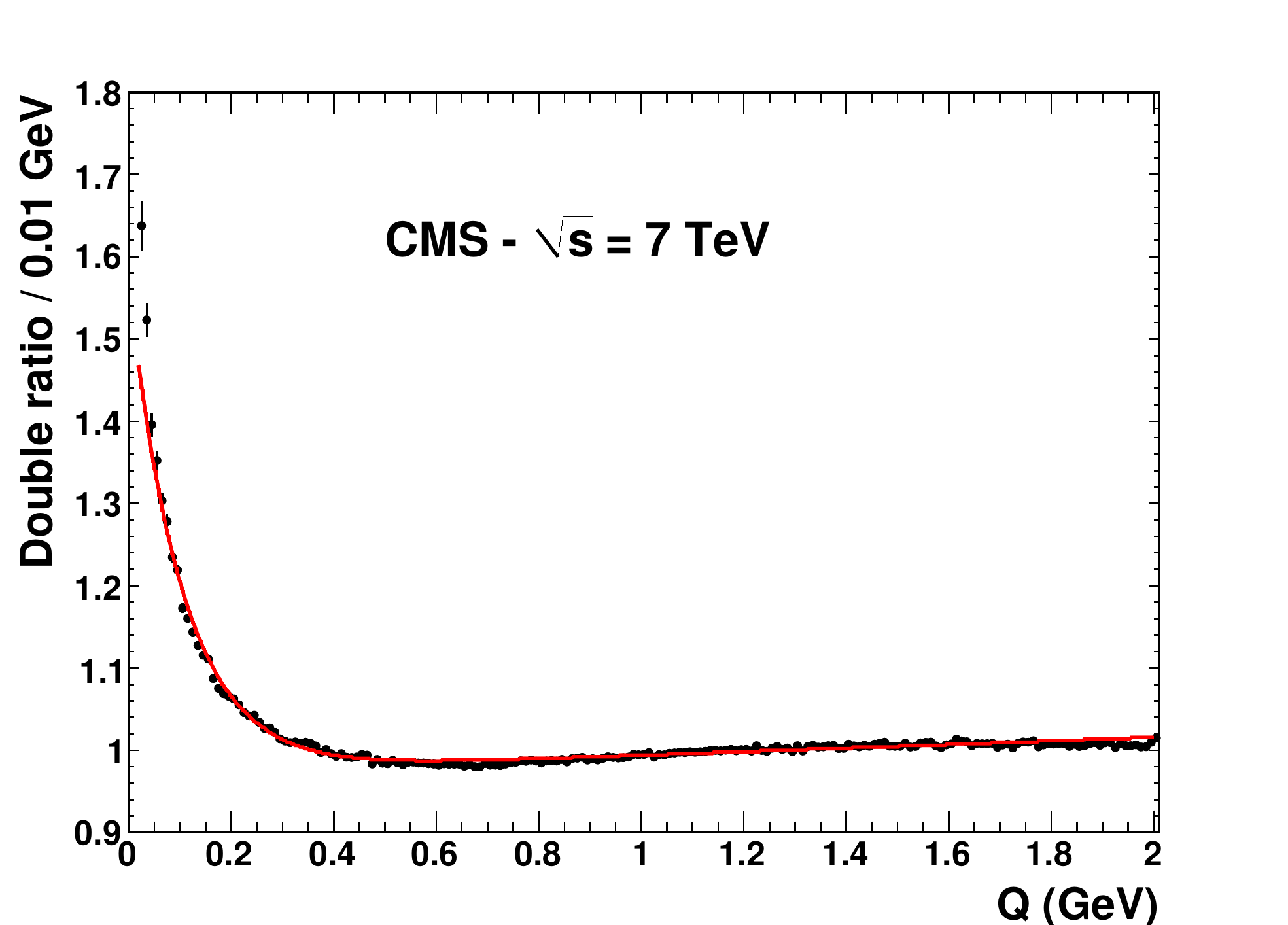}

        \caption{
Distribution, for $Q > 0.02$ GeV, of the double ratio ${\cal{R}}$
defined in Eq.~(\ref{eq:doubleratio}),
for data at $\sqrt{s} = 0.9$ (left)  and 7 TeV (right).
The reference sample is obtained from same-sign charged particles from mixed events with similar multiplicities,
and the MC simulation is {\sc PYTHIA6.4} tune Z2.
	The lines are the results
of fits using the exponential parametrization for $\Omega(Qr)$,
 with the values of the parameters given in the text.
The error bars on the data points are statistical only.
}
        \label{f:dRZ2}
    \end{center}
\end{figure}

Figure~\ref{f:dRZ2}
shows the distributions of the double ratio  ${\cal{R}}$
for $Q > 0.02$ GeV and both centre-of-mass energies,
computed using the tune Z2 of the {\sc PYTHIA6.422} simulation, which best describes the measured track distributions
(in particular the charged-particle multiplicity).
The shapes fitted with the exponential parameterization
$\Omega(Qr)= e^{-Qr}$ in Eq.~(\ref{eq:exp})
are superimposed and the results of the fits are given in Table~\ref{t:resultsDR}.
The values of the two parameters $r$ and  $\lambda$ are basically related  to different features of the distributions:
the width of the peak at small $Q$ to $r$ and the height to $\lambda$;
they are, however, strongly correlated,
 with correlation coefficients of about $86\%$.
The fit quality is poor, as can be seen from the values of $\chi^2/N_{{\rm dof}}$.
Gaussian parameterizations,
which are used by some experiments,
provide values of $\chi^2/N_{{\rm dof}}$ larger than 9, which confirms the observation in~\cite{bec_prl}
that an exponential parameterization is preferred.
Compared to an exponential shape, alternative functions, as defined in \cite{kozlov,Biyajima}, and the L\'evy parameterization,
$\Omega(Qr)= e^{-(Qr)^\alpha}$~\cite{levy}, yield fits of only slightly better quality.
However, the latter better describes the data at low $Q$, with values of $\alpha <1$ as in~\cite{bec_prl}.
As a cross-check of the stability of the measurement of the width of the peak at small $Q$
and of the fact that it does not depend on the fit quality, the average values
(first moment) of the $\Omega(Qr)$ distributions
over the same interval in $Q$ are found to be consistent for the different functions.
More discussion on the shape of the ${\cal{R}}$ distribution and on the fit quality can be found at the end
of this section.
\begin{table}[hbt]
\begin{center}
\caption{
Results of
of fits using the exponential parametrization for $\Omega(Qr)$
to  the double ratio  ${\cal R}$, at $\sqrt{s} = 0.9$ and 7 TeV.
The reference sample is obtained from same-sign charged particles from mixed events
with similar multiplicities, and the MC simulation is {\sc PYTHIA6.4} tune Z2.
 Errors are statistical only.
 \label{t:resultsDR}
}
\begin{tabular}{l|c|c|c|c|c}
\hline
		$\sqrt{s}$	&	 $\chi^2/N_{{\rm dof}}$ & C & $\lambda$ & r~(fm)  & $\delta$ $(10^{-2}$ GeV$^{-1})$  \\

\hline

\hline
$0.9$ TeV &				 $2.5$  & $0.965\pm 0.001$ & $0.616\pm 0.011$ & $1.56\pm 0.02$ & $2.8\pm 0.1$ \\
\hline
$7$ TeV &				 $3.8$  & $0.971\pm 0.001$ & $0.618\pm 0.009$ & $1.89\pm 0.02$ & $2.2\pm 0.1$ \\
\hline

\end{tabular}
\end{center}
\end{table}
As discussed in~\cite{bec_prl}, the main experimental uncertainty is due to the choice
of the reference sample.
In addition, we consider here the systematic
uncertainty due to the choice of the MC sample;
it
is obtained from the r.m.s.~spread
of  results  obtained using the various MC simulations listed in Sect.~\ref{s:datasets}.
The results at 7 TeV show a larger dependence on the MC choice than at 0.9 TeV.
Finally, the uncertainty related to the Coulomb corrections is taken to be  $\pm 15\%$,
as determined in \cite{bec_prl}, which covers the spread  from  the different parameterizations
\cite{bowler}.
The Coulomb corrections affect the signal mainly at very low $Q$, leading to an uncertainty of
 $\pm 2.8\%$ on $\lambda$ and  $\pm 0.8\%$ on $r$.
Contributions to the systematic uncertainties are reported in Table~\ref{t:sys},
and the total systematic uncertainties are obtained from their quadratic sum.
It was checked that reducing the fit range to
$0.04 < Q < 2 \GeV$, thus excluding the first two points at low $Q$ in Fig.~\ref{f:dRZ2},
gives consistent results within errors.
\begin{table}[hbt]
\begin{center}
\caption{ Systematic
uncertainties on the parameters $\lambda$ and $r$ at two different
centre-of-mass energies.
\label{t:sys}
}
\begin{tabular}{l|c|c|c|c}
\hline
  $\sqrt{s}$ 	      &\multicolumn{2} {c|} {0.9 TeV}& \multicolumn{2} {c} {7 TeV}\\
\hline
                      & $\lambda$ & $r$ (fm) & $\lambda$ & $r$ (fm)  \\
\hline
Choice of the reference sample  & 0.017& 0.11 & 0.015& 0.10 \\
Choice of MC dataset   	       	& 0.009& 0.05 & 0.032& 0.16 \\	
Effect of Coulomb corrections  	& 0.017& 0.01 & 0.017& 0.02 \\
\hline
Total  			    	& 0.026& 0.12& 0.039& 0.19  \\
\hline
\end{tabular}
\end{center}
\end{table}
The BEC parameters are thus measured to be
\begin{eqnarray*}
r&=&1.56\pm 0.02\, (\rm{stat.}) \pm\, 0.12\, (\rm{syst.})\, \rm{ fm} \\
 \lambda&=&0.616\pm 0.011\, (\rm{stat.}) \pm\, 0.026\, (\rm{syst.})
\end{eqnarray*}
at $\sqrt{s}=0.9$~TeV and
\begin{eqnarray*}
r&=&1.89\pm 0.02\, (\rm{stat.}) \pm\, 0.19\,(\rm{syst.})\, \rm{ fm} \\
\lambda&=&0.618\pm 0.009\, (\rm{stat.}) \pm\, 0.039\, (\rm{syst.})
\end{eqnarray*}
at $\sqrt{s}=7$~TeV.

\begin{figure}[!h]
    \begin{center}
        \includegraphics[width=0.975\textwidth, angle=90]{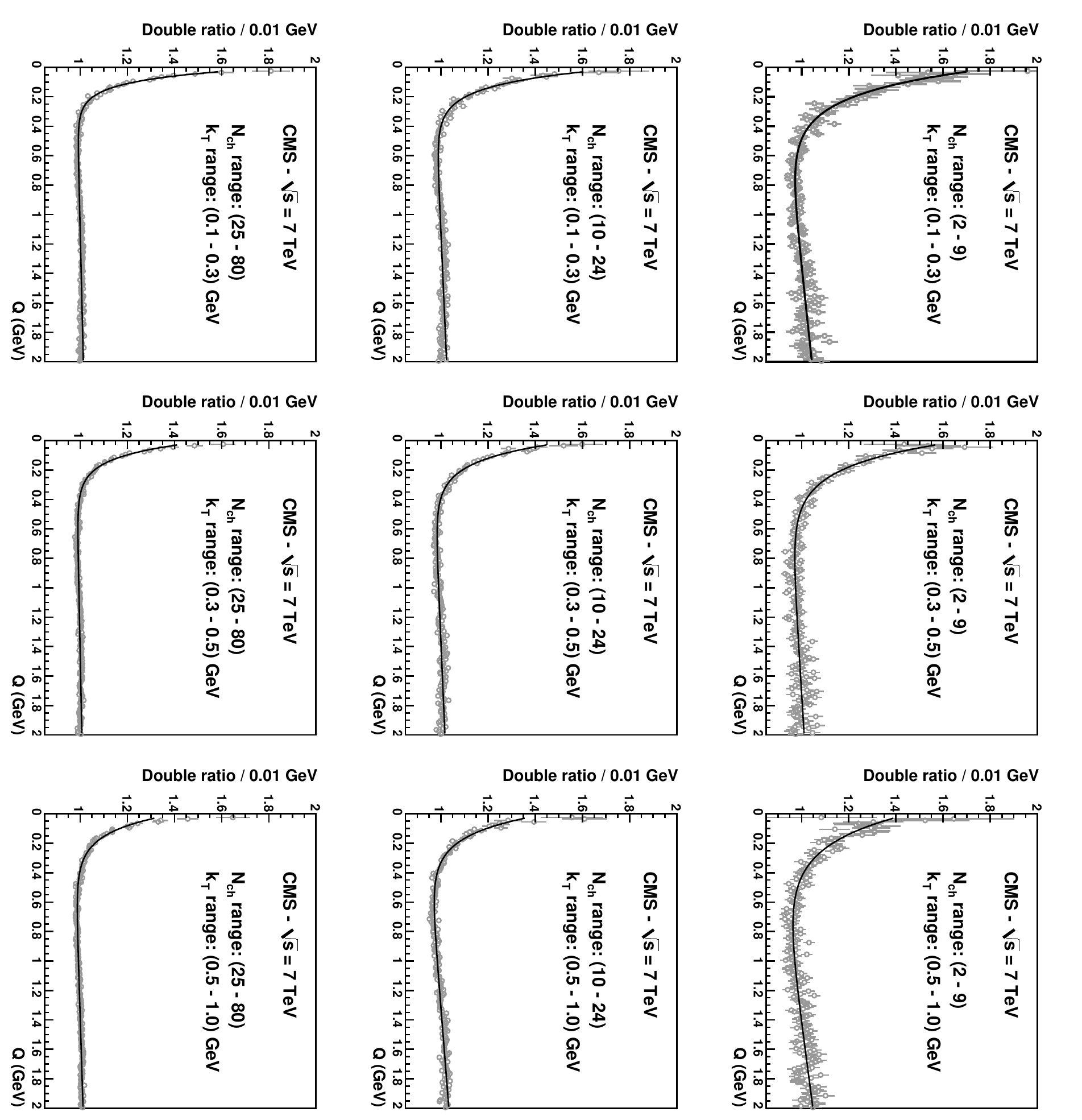}
        \caption{Distributions of the double ratio ${\cal{R}}$ as a function of $Q$,
         for three intervals in $k_{\mathrm{T}}$ and three intervals in charged-particle multiplicity
         in the event, $N_{\mathrm{ch}}$, for $\sqrt{s}=$ 7 TeV.
The lines are the results of
fits using the exponential parametrization for $\Omega(Qr)$,
with the values of the parameters given in Table~\ref{t:rvskt}.
The error bars are statistical only.}
        \label{f:kt_nch}
    \end{center}
\end{figure}

\begin{figure}[!h]
    \begin{center}
        \includegraphics[height=0.57\textwidth, angle=0]{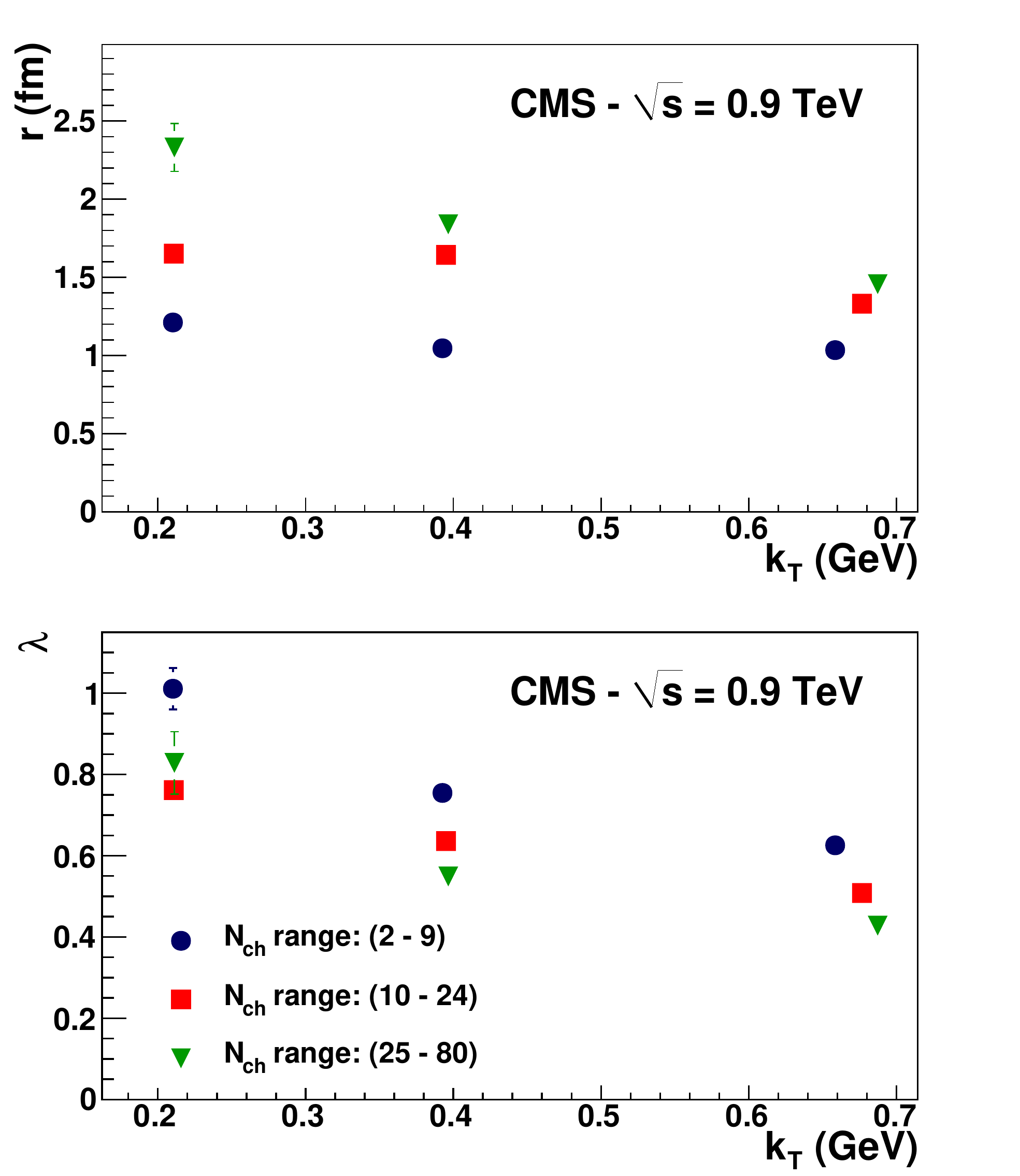}
        \includegraphics[height=0.57\textwidth, angle=0]{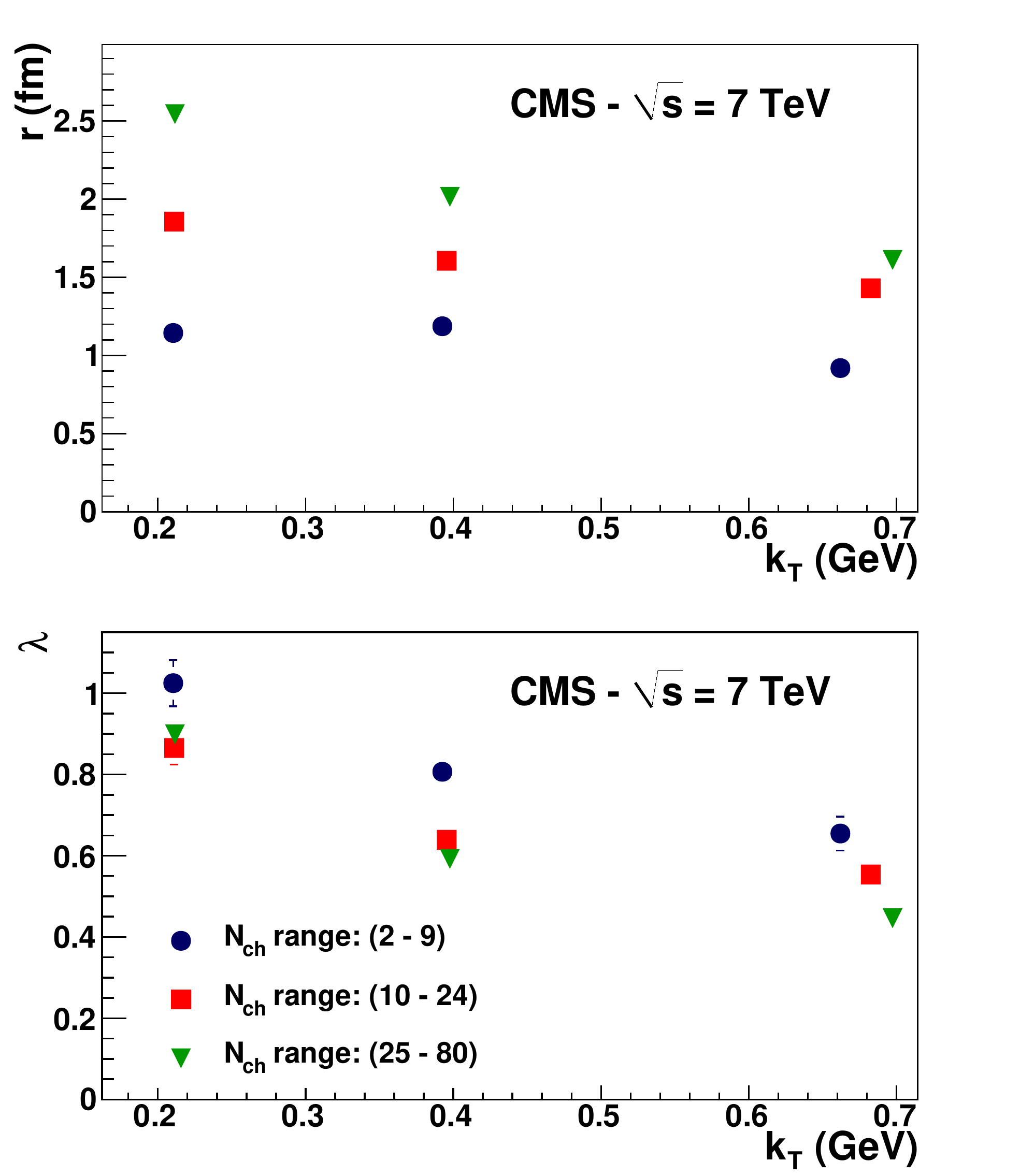}
        \caption{Values of the parameters  $r$ (top) and  $\lambda$ (bottom), as a function of $k_{\mathrm{T}}$
 in three intervals of charged-particle multiplicity in the event, $N_{\mathrm{ch}}$, for $\sqrt{s}= 0.9$ (left)
and 7 TeV (right). The points are presented at the position corresponding to the mean value of %the variable
$k_{\mathrm{T}}$ in the considered interval of $N_{\mathrm{ch}}$.
The error bars are  statistical only (in some cases they are smaller than the marker size).
The systematic uncertainties
are discussed in the text.
}
        \label{f:r_vs_kt_nch}
    \end{center}
\end{figure}

\begin{figure}[hbt]
   \begin{center}
       \includegraphics[width=0.45\textwidth, angle=90]{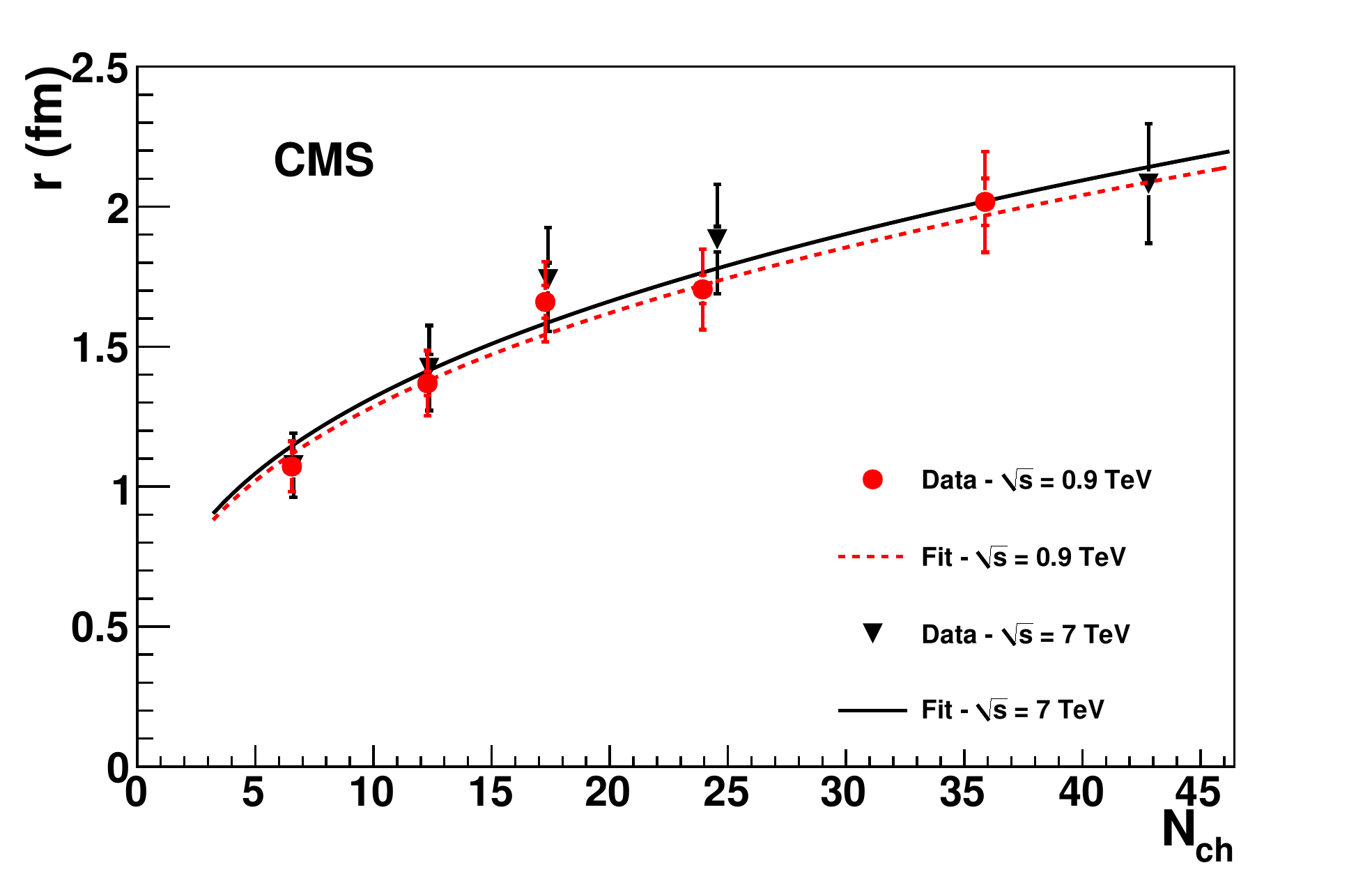}
       \caption{
Parameter $r$ as a function
of the charged-particle multiplicity in the event, $N_{\mathrm{ch}}$, for $\sqrt{s} = 0.9$ and 7 TeV.
The dotted and
solid lines represent the results of the fits described in the text to the 0.9  and 7 TeV data, respectively.
The inner error bars represent the statistical uncertainties  and the outer error bars the statistical
and systematic uncertainties,
added in quadrature. The systematic uncertainties are dominating and are point-to-point correlated.
}
       \label{f:nch_dep_7tev_900gev}
   \end{center}
\end{figure}

As will be shown below, the increase of $r$ is related to the different
average charged-particle multiplicities at the two energies, while the value of the $\lambda$ parameter is stable.
The BEC signal is studied as a function of  the charged-particle multiplicity in the event, $N_{\mathrm{ch}}$, as in~\cite{bec_prl},
and of the pair average transverse momentum $k_{\mathrm{T}}$, defined as half of the absolute
vector sum of the two transverse momenta, $k_{\mathrm{T}} = |{\bf p}_{\mathrm{T},1}+{\bf p}_{\mathrm{T},2}|/2$.
A dependence on $k_{\mathrm{T}}$  has been observed at the Tevatron \cite{E735} and at RHIC \cite{rich},
where it is associated with the system collective expansion.
Figure~\ref{f:kt_nch} shows the double ratio ${\cal{R}}$  as a function of $Q$ for different values of
$N_{\mathrm{ch}}$ and $k_{\mathrm{T}}$.
The $k_{\mathrm{T}}$ dependence of the  $r$  and $\lambda$  parameters  for  three intervals of multiplicity,
obtained with the exponential parameterization,
is  shown in  Fig.~\ref{f:r_vs_kt_nch} and given in Table~\ref{t:rvskt} for the 0.9  and 7 TeV data.
The effective radius $r$  is observed to increase with multiplicity,
for all reference samples and for all MC models and tunes, in agreement with previous results.
 At low multiplicity, $r$ is approximately independent of  $k_{\mathrm{T}}$, and it decreases with $k_{\mathrm{T}}$
as $N_{\mathrm{ch}}$ increases.
The $\lambda$ parameter decreases with increasing multiplicity and  $k_{\mathrm{T}}$.
The systematic uncertainties are estimated  to be the same as for the overall measurements
($4.2\%$ and  $6.3\%$  for $\lambda$,  $7.9\%$ and $10.1\%$ for $r$, at 0.9
and 7 TeV, respectively). It should be noted that these uncertainties are point-to-point correlated,
since the effects of the choice of the various reference samples and of the various
MC simulations are very similar for the different subsamples.
The 0.9 TeV results agree with those of ALICE~\cite{ALICE_BEC}.
\begin{table}[!h]
\begin{center}
\caption{
Results
of fits using the exponential parametrization for $\Omega(Qr)$
 to the double ratios  ${\cal R}$, for three intervals in  $k_{\mathrm{T}}$
and three intervals in charged-particle multiplicity in the event, $N_{\mathrm{ch}}$, for $ \sqrt{s} = 0.9$ and 7 TeV.
The errors are statistical only. The systematic uncertainties, which are point-to-point correlated,
are estimated from the relative uncertainties affecting the overall measurements (see text).
\label{t:rvskt}
}
{\scriptsize
\begin{tabular}{c|c|c|c|c|c|c}
\hline
$k_{\mathrm{T}}$  (GeV)      & $N_{\mathrm{ch}}$   &       $\chi^2/N_{{\rm dof}}$    &         $C$    &  $\lambda$     & $r$ (fm)       &  $\delta$ $(10^{-2}$ GeV$^{-1}$)  \\
\hline
\multicolumn{7}{c}{$\sqrt{s}= 0.9$ TeV }\\
\hline

0.10 - 0.30 &   2 -  9  &  $1.1$  & $0.925\pm 0.006$ & $1.011\pm 0.051$ & $1.211\pm 0.057$ & $6.1\pm 0.6$  \\
0.10 - 0.30 &  10 -  24 &  $1.5$  & $0.969\pm 0.002$ & $0.761\pm 0.034$ & $1.652\pm 0.057$ & $2.9\pm 0.2$  \\
0.10 - 0.30 &  25 -  80 &  $1.1$  & $0.984\pm 0.002$ & $0.828\pm 0.077$ & $2.331\pm 0.153$ & $1.6\pm 0.2$  \\
0.30 - 0.50 &   2 -  9  &  $1.1$  & $0.912\pm 0.007$ & $0.754\pm 0.027$ & $1.046\pm 0.049$ & $6.0\pm 0.6$  \\
0.30 - 0.50 &  10 -  24 &  $1.3$  & $0.970\pm 0.002$ & $0.636\pm 0.023$ & $1.643\pm 0.051$ & $2.3\pm 0.2$  \\
0.30 - 0.50 &  25 -  80 &  $1.2$  & $0.984\pm 0.002$ & $0.549\pm 0.033$ & $1.839\pm 0.089$ & $1.2\pm 0.2$  \\
0.50 - 1.00 &   2 -  9  &  $1.2$  & $0.911\pm 0.009$ & $0.626\pm 0.039$ & $1.034\pm 0.079$ & $6.6\pm 0.8$  \\
0.50 - 1.00 &  10 -  24 &  $1.1$  & $0.957\pm 0.003$ & $0.508\pm 0.024$ & $1.331\pm 0.059$ & $3.4\pm 0.2$  \\
0.50 - 1.00 &  25 -  80 &  $1.1$  & $0.979\pm 0.003$ & $0.428\pm 0.029$ & $1.456\pm 0.086$ & $1.5\pm 0.2$  \\
\hline
\multicolumn{7}{c}{$\sqrt{s}= 7$ TeV }\\
\hline

0.10 - 0.30 &   2 -  9  &  $1.1$  & $0.910\pm 0.008$ & $1.025\pm 0.057$ & $1.144\pm 0.062$ & $7.3\pm 0.7$  \\
0.10 - 0.30 &  10 -  24 &  $1.5$  & $0.970\pm 0.002$ & $0.865\pm 0.041$ & $1.856\pm 0.065$ & $2.8\pm 0.2$  \\
0.10 - 0.30 &  25 -  80 &  $1.5$  & $0.984\pm 0.001$ & $0.899\pm 0.039$ & $2.544\pm 0.076$ & $1.5\pm 0.1$  \\
0.30 - 0.50 &   2 -  9  &  $1.0$  & $0.935\pm 0.008$ & $0.807\pm 0.039$ & $1.187\pm 0.066$ & $4.1\pm 0.7$  \\
0.30 - 0.50 &  10 -  24 &  $1.5$  & $0.964\pm 0.002$ & $0.639\pm 0.023$ & $1.606\pm 0.050$ & $2.8\pm 0.2$  \\
0.30 - 0.50 &  25 -  80 &  $1.7$  & $0.982\pm 0.001$ & $0.592\pm 0.018$ & $2.015\pm 0.048$ & $1.3\pm 0.1$  \\
0.50 - 1.00 &   2 -  9  &  $0.9$  & $0.883\pm 0.013$ & $0.655\pm 0.042$ & $0.919\pm 0.078$ & $9.4\pm 1.1$  \\
0.50 - 1.00 &  10 -  24 &  $1.4$  & $0.936\pm 0.003$ & $0.554\pm 0.026$ & $1.430\pm 0.057$ & $5.2\pm 0.2$  \\
0.50 - 1.00 &  25 -  80 &  $1.8$  & $0.973\pm 0.001$ & $0.446\pm 0.016$ & $1.611\pm 0.048$ & $2.0\pm 0.1$  \\

\hline
\end{tabular}
}
\end{center}
\end{table}
Figure~\ref{f:nch_dep_7tev_900gev} presents the
distribution of the parameter $r$ as a function of
$N_{\mathrm{ch}}$
for both centre-of-mass energies.
The measurements are consistent, indicating that the difference between the values of $r$ obtained for
the two global samples are accounted for by the different average charged-particle multiplicities,
which are 12.1 and 19.2 for the 0.9 and 7 TeV cases, respectively.
This trend is consistent with the result of a similar comparison
between data at 0.9 and 2.36 TeV~\cite{bec_prl}.
The multiplicity dependence is fitted as
$r(N_{\mathrm{ch}}) = a\cdot N_{\mathrm{ch}}^{1/3}$ \cite{N13}, giving
$a=0.597\pm 0.009$~(stat.)~$\pm\, 0.047$~(syst.)~fm at 0.9 TeV and
$a=0.612\pm0.007$~(stat.)~$\pm\, 0.063$~(syst.)~fm at 7 TeV.

\begin{figure}[hbt]
    \begin{center}
 		\includegraphics[height=0.45\textwidth,angle=90]{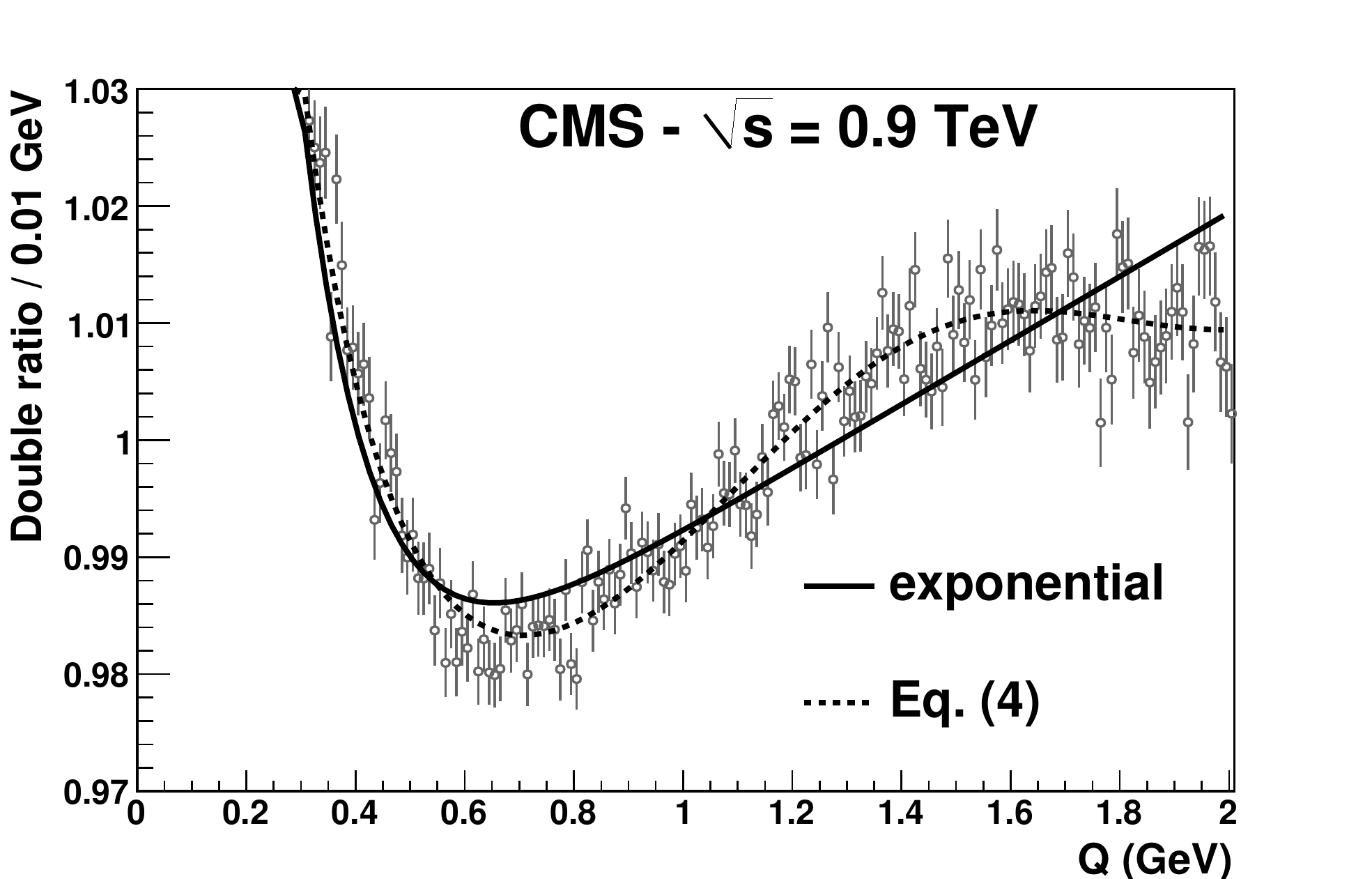}
                \includegraphics[height=0.45\textwidth,angle=90]{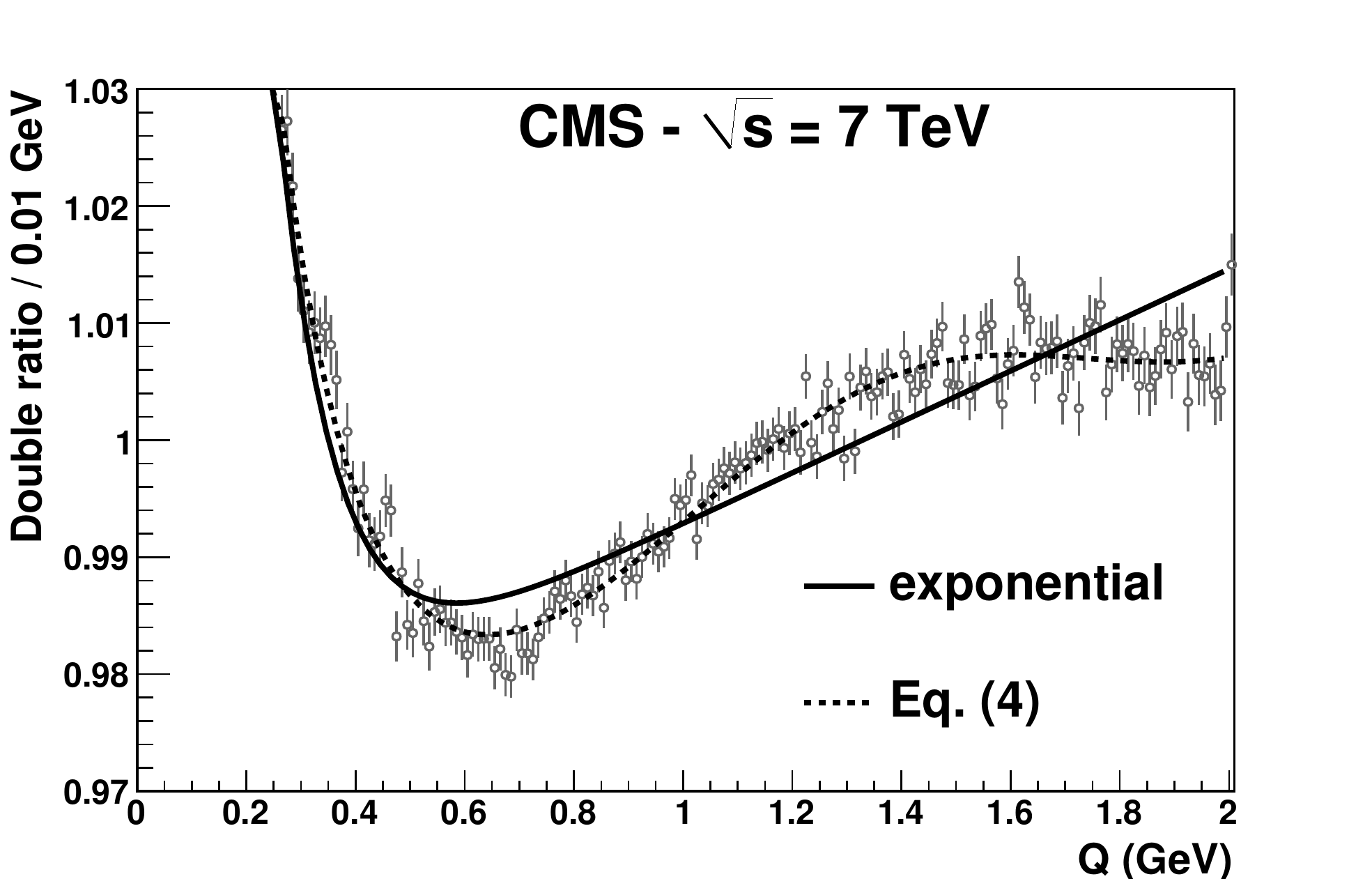}

        \caption{Detail of the distribution of the double ratio ${\cal{R}}$ for $\sqrt{s} = 0.9$ (left) and
        7 TeV (right). The dotted lines correspond to fits with Eq.~(\ref{eq:alevy}), and
	the %dashed
  solid lines to exponential fits.
 Note the enlarged scale on the y axis. The  error bars are statistical only. }

        \label{f:7TeV_zoom}
    \end{center}
\end{figure}

\begin{figure}[!h]
    \begin{center}
	\includegraphics[height=0.9\textwidth, angle=90]{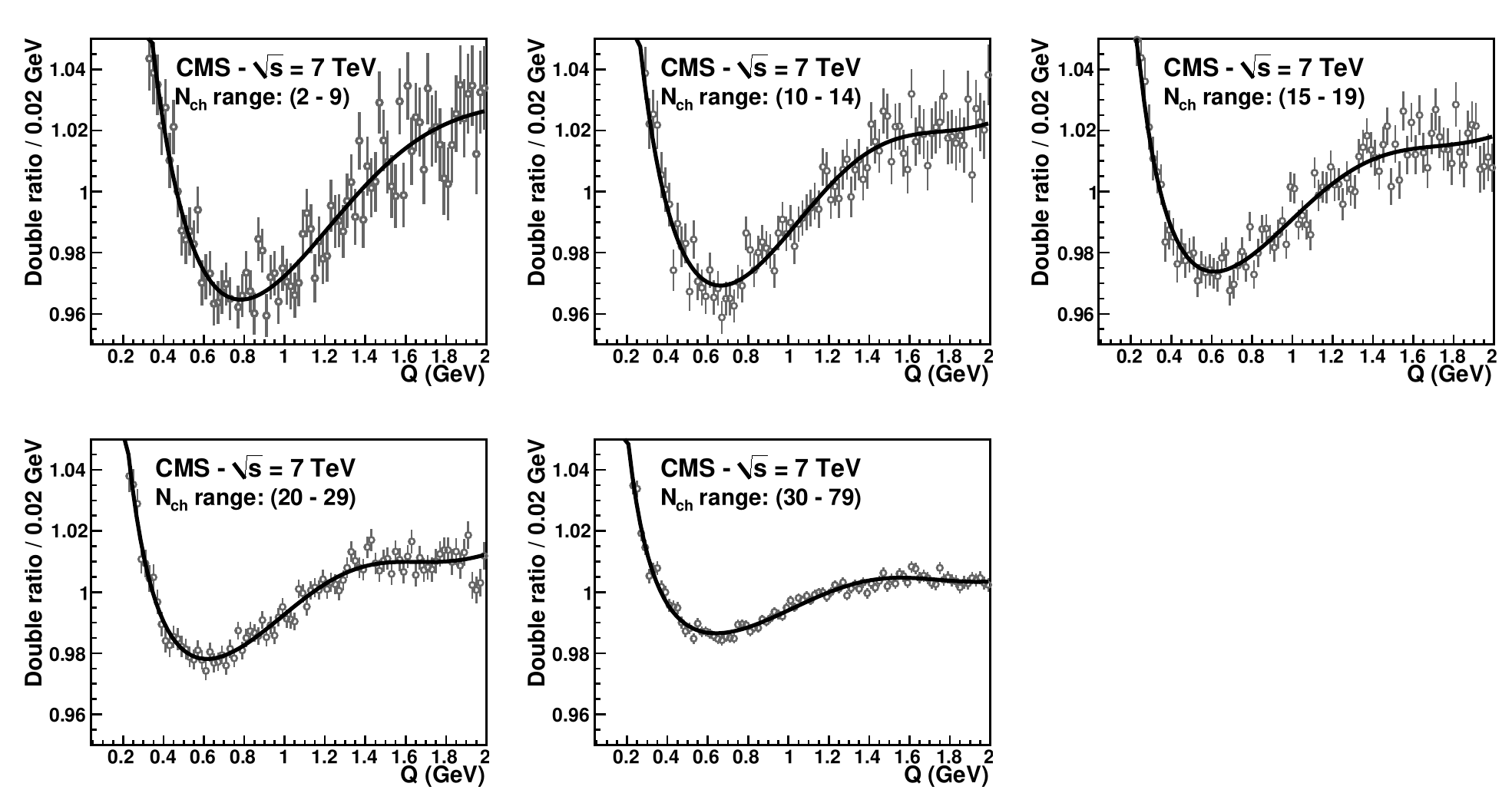}
      \caption{
Detail of the distribution of the double ratio ${\cal{R}}$
        for $\sqrt{s} = 7$ TeV %for
        using different intervals of charged-particle multiplicity in the event ($N_{\mathrm{ch}}$).
The  lines are  fits to the data with Eq.~(\ref{eq:alevy}). The error bars are statistical only. }
        \label{f:dip}
    \end{center}
\end{figure}

\begin{figure}[!h]
    \begin{center}
	\includegraphics[height=0.6\textwidth, angle=90]{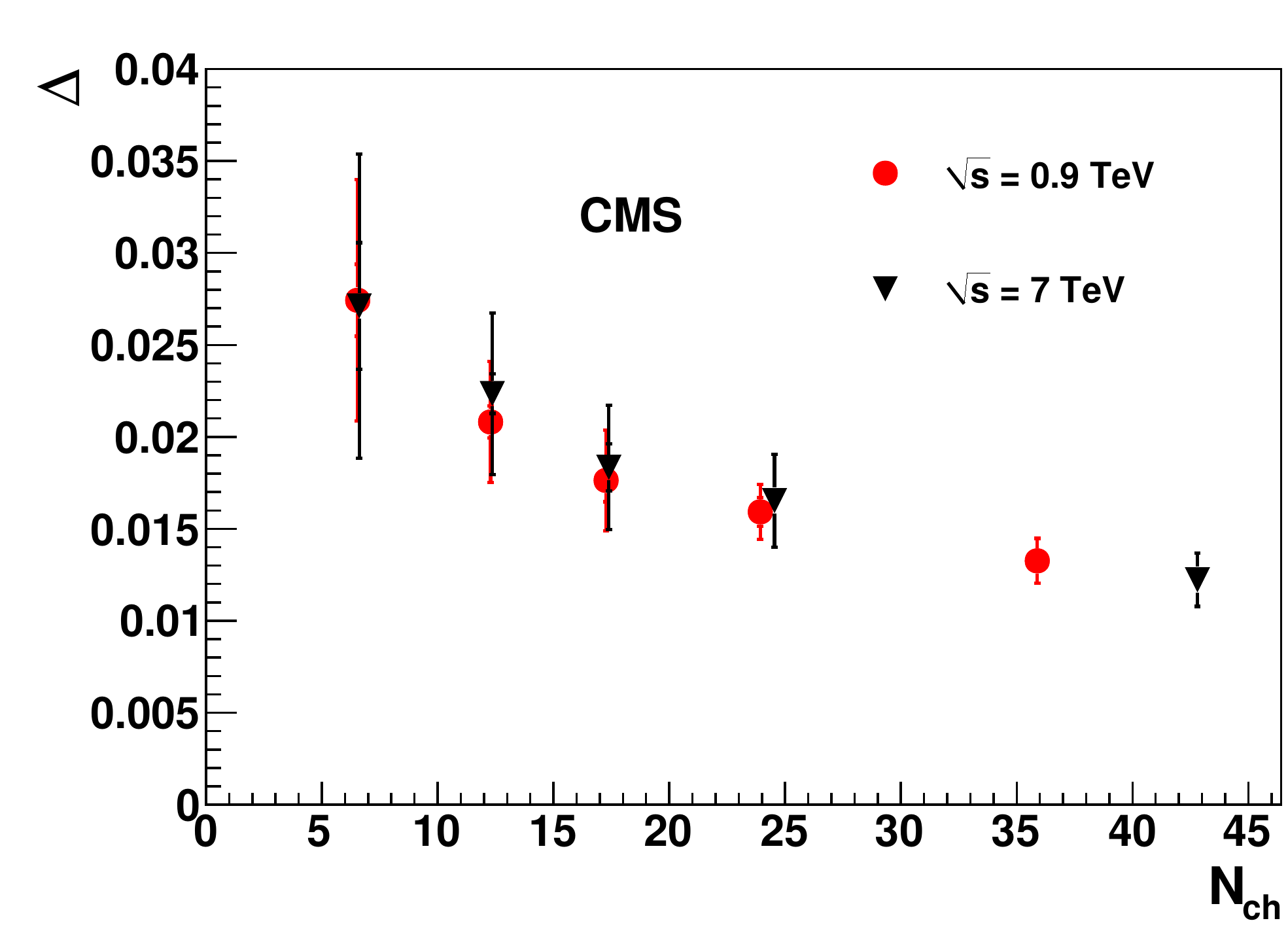}
      \caption{Depth $\Delta$ of
the dip in
the anticorrelation region,
as a function of the charged-particle multiplicity in the event,
for $\sqrt{s} = 0.9$ and 7 TeV.
The inner error bars represent the statistical errors and the outer error bars the statistical and systematic errors,
added in quadrature. The systematic uncertainties are dominating and are point-to-point correlated.}
        \label{f:dip_trend}
    \end{center}
\end{figure}

As
was noted above and
can be deduced from the $\chi^2/N_{{\rm dof}}$ in Table~\ref{t:resultsDR},
none of the quoted functions is able to provide a good description of
 the  ${\cal{R}}$  distributions.
This is due to an anticorrelation effect between same-sign charged particles for $Q$ values just above the signal region
(dip with  ${\cal{R}}< 1$), as shown in Fig.~\ref{f:7TeV_zoom}.
This anticorrelation is
observed in the double ratio at both energies with any choice of reference
sample and MC simulation.
It shows little
sensitivity to $k_{\mathrm{T}}$, while it decreases with increasing
charged-particle multiplicity in the event, as shown  in Fig.~\ref{f:dip} for the 7 TeV data.
This detailed observation
is made possible by the large data samples studied here, and constitutes the first evidence
 of this effect at the LHC. Such a structure was observed in $e^+e^-$ collisions at LEP~\cite{l3_dip}.
The presence of a region of anticorrelation between same-sign charged particles has been explained
in~\cite{alevy}.
In this reference,
a parameterization for $R(Q)$
has been proposed, aimed at describing the time evolution of the source:
\begin{linenomath}
\begin{equation}
\label{eq:alevy}
 R(Q) = C\left[1+ \lambda({\text{cos}}\left[ (r_0 Q)^2+\text{tan}(\alpha\pi/4)( Qr)^{\alpha}\right] e^{-(Qr)^\alpha}) \right] \cdot(1+\delta Q).
\end{equation}
\end{linenomath}
\noindent
Here $r_0$ is related to the proper time of the onset of  particle production, and $r$ enters
in both the exponential and the oscillation factors.
As visible in Fig.~\ref{f:7TeV_zoom},  fits with  Eq.~(\ref{eq:alevy})
are of good quality, much better than with an exponential function,
and the $\chi^2/N_{{\rm dof}}$
is 1.1 both for the 0.9  and  7 TeV data samples.
The depth of the dip in the anticorrelation region is
measured as the difference $\Delta$ between the baseline curve defined as
$ C \cdot (1+ \delta Q)$ and the value of $R$ defined by Eq.~(\ref{eq:alevy}) at its minimum.
Results are shown in Fig.~\ref{f:dip_trend}.
The depths are found to decrease with
$N_{\mathrm{ch}}$
consistently for  the two centre-of-mass energies.
The systematic errors have been computed from the r.m.s.~spread of the results obtained with the various
reference samples and MC simulations. It has been checked that these results are robust:
when the fitting range is extended to $Q$ = 5 GeV the results are consistent within errors
and the trend is similar.
\section{Conclusions \label{s:conclusions}}

Bose--Einstein correlations have been measured using data collected with the CMS experiment
in proton-proton collisions at the LHC, with centre-of-mass energies of 0.9 and 7 TeV.
The signal is observed as an enhancement of pairs of same-sign charged particles with small relative momentum.
The parameters are obtained from
fits using the exponential parametrization for $\Omega(Qr)$
to the distribution of $Q$.
In agreement with previous results, an increase of the effective emission radius $r$
with charged-particle multiplicity in the event is observed, which accounts for the increase of $r$ from 0.9 to 7 TeV.
The parameter $r$ is nearly independent of the average transverse momentum of the pair of particles,
and decreases with $k_{\mathrm{T}}$  in events with large charged-particle multiplicity. Anticorrelations between same-sign charged particles
are observed for $Q$ values above the signal region as previously reported with LEP data.
The anticorrelation effects decrease with increasing charged-particle multiplicity in the event.
\section*{Acknowledgements \label{s:ack}}
We wish to congratulate our colleagues in the CERN accelerator departments for
the excellent performance of the LHC machine. We thank the technical and
administrative staff at CERN and other CMS institutes, and acknowledge support
from: FMSR (Austria); FNRS and FWO (Belgium); CNPq, CAPES, FAPERJ, and FAPESP
(Brazil); MES (Bulgaria); CERN; CAS, MoST, and NSFC (China); COLCIENCIAS
(Colombia); MSES (Croatia); RPF (Cyprus); Academy of Sciences and NICPB
(Estonia); Academy of Finland, ME, and HIP (Finland); CEA and CNRS/IN2P3
(France); BMBF, DFG, and HGF (Germany); GSRT (Greece); OTKA and NKTH (Hungary);
DAE and DST (India); IPM (Iran); SFI (Ireland); INFN (Italy); NRF (Korea); LAS
(Lithuania); CINVESTAV, CONACYT, SEP, and UASLP-FAI (Mexico); PAEC (Pakistan);
SCSR (Poland); FCT (Portugal); JINR (Armenia, Belarus, Georgia, Ukraine,
    Uzbekistan); MST and MAE (Russia); MSTDS (Serbia); MICINN and CPAN (Spain);
Swiss Funding Agencies (Switzerland); NSC (Taipei); TUBITAK and TAEK (Turkey);
STFC (United Kingdom); DOE and NSF (USA).
\bibliography{auto_generated}   % will be created by the tdr script.

\cleardoublepage\appendix\section{The CMS Collaboration \label{app:collab}}\begin{sloppypar}\hyphenpenalty=5000\widowpenalty=500\clubpenalty=5000\input{QCD-10-023-authorlist.tex}\end{sloppypar}
\end{document}

%% file: ptdr-definitions.tex
%%%%%%%%%%%%%%%%%%%%%%%%%%%%%%%%%%%%%%%%%%%%%%%%%%%%%%%%%%%%%%%%%%%%
%
%  Common definitions
%
%  N.B. use of \providecommand rather than \newcommand means
%       that a definition is ignored if already specified
%
%                                              L. Taylor 18 Feb 2005
%%%%%%%%%%%%%%%%%%%%%%%%%%%%%%%%%%%%%%%%%%%%%%%%%%%%%%%%%%%%%%%%%%%%

% Some shorthand
% turn off italics
\providecommand {\etal}{\mbox{et al.}\xspace} %et al. - no preceding comma
\providecommand {\ie}{\mbox{i.e.}\xspace}     %i.e.
\providecommand {\eg}{\mbox{e.g.}\xspace}     %e.g.
\providecommand {\etc}{\mbox{etc.}\xspace}     %etc.
\providecommand {\vs}{\mbox{\sl vs.}\xspace}      %vs.
\providecommand {\mdash}{\ensuremath{\mathrm{-}}} % for use within formulas

% some terms whose definition we may change
\providecommand {\Lone}{Level-1\xspace} % Level-1 or L1 ?
\providecommand {\Ltwo}{Level-2\xspace}
\providecommand {\Lthree}{Level-3\xspace}

% Some software programs (alphabetized)
\providecommand{\ACERMC} {\textsc{AcerMC}\xspace}
\providecommand{\ALPGEN} {{\textsc{alpgen}}\xspace}
\providecommand{\CHARYBDIS} {{\textsc{charybdis}}\xspace}
\providecommand{\CMKIN} {\textsc{cmkin}\xspace}
\providecommand{\CMSIM} {{\textsc{cmsim}}\xspace}
\providecommand{\CMSSW} {{\textsc{cmssw}}\xspace}
\providecommand{\COBRA} {{\textsc{cobra}}\xspace}
\providecommand{\COCOA} {{\textsc{cocoa}}\xspace}
\providecommand{\COMPHEP} {\textsc{CompHEP}\xspace}
\providecommand{\EVTGEN} {{\textsc{evtgen}}\xspace}
\providecommand{\FAMOS} {{\textsc{famos}}\xspace}
\providecommand{\GARCON} {\textsc{garcon}\xspace}
\providecommand{\GARFIELD} {{\textsc{garfield}}\xspace}
\providecommand{\GEANE} {{\textsc{geane}}\xspace}
\providecommand{\GEANTfour} {{\textsc{geant4}}\xspace}
\providecommand{\GEANTthree} {{\textsc{geant3}}\xspace}
\providecommand{\GEANT} {{\textsc{geant}}\xspace}
\providecommand{\HDECAY} {\textsc{hdecay}\xspace}
\providecommand{\HERWIG} {{\textsc{herwig}}\xspace}
\providecommand{\HIGLU} {{\textsc{higlu}}\xspace}
\providecommand{\HIJING} {{\textsc{hijing}}\xspace}
\providecommand{\IGUANA} {\textsc{iguana}\xspace}
\providecommand{\ISAJET} {{\textsc{isajet}}\xspace}
\providecommand{\ISAPYTHIA} {{\textsc{isapythia}}\xspace}
\providecommand{\ISASUGRA} {{\textsc{isasugra}}\xspace}
\providecommand{\ISASUSY} {{\textsc{isasusy}}\xspace}
\providecommand{\ISAWIG} {{\textsc{isawig}}\xspace}
\providecommand{\MADGRAPH} {\textsc{MadGraph}\xspace}
\providecommand{\MCATNLO} {\textsc{mc@nlo}\xspace}
\providecommand{\MCFM} {\textsc{mcfm}\xspace}
\providecommand{\MILLEPEDE} {{\textsc{millepede}}\xspace}
\providecommand{\ORCA} {{\textsc{orca}}\xspace}
\providecommand{\OSCAR} {{\textsc{oscar}}\xspace}
\providecommand{\PHOTOS} {\textsc{photos}\xspace}
\providecommand{\PROSPINO} {\textsc{prospino}\xspace}
\providecommand{\PYTHIA} {{\textsc{pythia}}\xspace}
\providecommand{\SHERPA} {{\textsc{sherpa}}\xspace}
\providecommand{\TAUOLA} {\textsc{tauola}\xspace}
\providecommand{\TOPREX} {\textsc{TopReX}\xspace}
\providecommand{\XDAQ} {{\textsc{xdaq}}\xspace}

%  Experiments
\providecommand {\DZERO}{D\O\xspace}     %etc.

% Measurements and units...

\providecommand{\de}{\ensuremath{^\circ}}
\providecommand{\ten}[1]{\ensuremath{\times \text{10}^\text{#1}}}
\providecommand{\unit}[1]{\ensuremath{\text{\,#1}}\xspace}
\providecommand{\mum}{\ensuremath{\,\mu\text{m}}\xspace}
\providecommand{\micron}{\ensuremath{\,\mu\text{m}}\xspace}
\providecommand{\cm}{\ensuremath{\,\text{cm}}\xspace}
\providecommand{\mm}{\ensuremath{\,\text{mm}}\xspace}
\providecommand{\mus}{\ensuremath{\,\mu\text{s}}\xspace}
\providecommand{\keV}{\ensuremath{\,\text{ke\hspace{-.08em}V}}\xspace}
\providecommand{\MeV}{\ensuremath{\,\text{Me\hspace{-.08em}V}}\xspace}
\providecommand{\GeV}{\ensuremath{\,\text{Ge\hspace{-.08em}V}}\xspace}
\providecommand{\gev}{\GeV}
\providecommand{\TeV}{\ensuremath{\,\text{Te\hspace{-.08em}V}}\xspace}
\providecommand{\PeV}{\ensuremath{\,\text{Pe\hspace{-.08em}V}}\xspace}
\providecommand{\keVc}{\ensuremath{{\,\text{ke\hspace{-.08em}V\hspace{-0.16em}/\hspace{-0.08em}}c}}\xspace}
\providecommand{\MeVc}{\ensuremath{{\,\text{Me\hspace{-.08em}V\hspace{-0.16em}/\hspace{-0.08em}}c}}\xspace}
\providecommand{\GeVc}{\ensuremath{{\,\text{Ge\hspace{-.08em}V\hspace{-0.16em}/\hspace{-0.08em}}c}}\xspace}
\providecommand{\TeVc}{\ensuremath{{\,\text{Te\hspace{-.08em}V\hspace{-0.16em}/\hspace{-0.08em}}c}}\xspace}
\providecommand{\keVcc}{\ensuremath{{\,\text{ke\hspace{-.08em}V\hspace{-0.16em}/\hspace{-0.08em}}c^\text{2}}}\xspace}
\providecommand{\MeVcc}{\ensuremath{{\,\text{Me\hspace{-.08em}V\hspace{-0.16em}/\hspace{-0.08em}}c^\text{2}}}\xspace}
\providecommand{\GeVcc}{\ensuremath{{\,\text{Ge\hspace{-.08em}V\hspace{-0.16em}/\hspace{-0.08em}}c^\text{2}}}\xspace}
\providecommand{\TeVcc}{\ensuremath{{\,\text{Te\hspace{-.08em}V\hspace{-0.16em}/\hspace{-0.08em}}c^\text{2}}}\xspace}

\providecommand{\pbinv} {\mbox{\ensuremath{\,\text{pb}^\text{$-$1}}}\xspace}
\providecommand{\fbinv} {\mbox{\ensuremath{\,\text{fb}^\text{$-$1}}}\xspace}
\providecommand{\nbinv} {\mbox{\ensuremath{\,\text{nb}^\text{$-$1}}}\xspace}
\providecommand{\percms}{\ensuremath{\,\text{cm}^\text{$-$2}\,\text{s}^\text{$-$1}}\xspace}
\providecommand{\lumi}{\ensuremath{\mathcal{L}}\xspace}
\providecommand{\Lumi}{\ensuremath{\mathcal{L}}\xspace}%both upper and lower
%
% Need a convention here:
\providecommand{\LvLow}  {\ensuremath{\mathcal{L}=\text{10}^\text{32}\,\text{cm}^\text{$-$2}\,\text{s}^\text{$-$1}}\xspace}
\providecommand{\LLow}   {\ensuremath{\mathcal{L}=\text{10}^\text{33}\,\text{cm}^\text{$-$2}\,\text{s}^\text{$-$1}}\xspace}
\providecommand{\lowlumi}{\ensuremath{\mathcal{L}=\text{2}\times \text{10}^\text{33}\,\text{cm}^\text{$-$2}\,\text{s}^\text{$-$1}}\xspace}
\providecommand{\LMed}   {\ensuremath{\mathcal{L}=\text{2}\times \text{10}^\text{33}\,\text{cm}^\text{$-$2}\,\text{s}^\text{$-$1}}\xspace}
\providecommand{\LHigh}  {\ensuremath{\mathcal{L}=\text{10}^\text{34}\,\text{cm}^\text{$-$2}\,\text{s}^\text{$-$1}}\xspace}
\providecommand{\hilumi} {\ensuremath{\mathcal{L}=\text{10}^\text{34}\,\text{cm}^\text{$-$2}\,\text{s}^\text{$-$1}}\xspace}

% Physics symbols ...

\providecommand{\PT}{\ensuremath{p_{\mathrm{T}}}\xspace}
\providecommand{\pt}{\ensuremath{p_{\mathrm{T}}}\xspace}
\providecommand{\ET}{\ensuremath{E_{\mathrm{T}}}\xspace}
\providecommand{\HT}{\ensuremath{H_{\mathrm{T}}}\xspace}
\providecommand{\et}{\ensuremath{E_{\mathrm{T}}}\xspace}
\providecommand{\Em}{\ensuremath{E\hspace{-0.6em}/}\xspace}
\providecommand{\Pm}{\ensuremath{p\hspace{-0.5em}/}\xspace}
\providecommand{\PTm}{\ensuremath{{p}_\mathrm{T}\hspace{-1.02em}/}\xspace}
\providecommand{\PTslash}{\ensuremath{{p}_\mathrm{T}\hspace{-1.02em}/}\xspace}
\providecommand{\ETm}{\ensuremath{E_{\mathrm{T}}^{\text{miss}}}\xspace}
\providecommand{\MET}{\ETm}
\providecommand{\ETmiss}{\ETm}
\providecommand{\ETslash}{\ensuremath{E_{\mathrm{T}}\hspace{-1.1em}/}\xspace}
\providecommand{\VEtmiss}{\ensuremath{{\vec E}_{\mathrm{T}}^{\text{miss}}}\xspace}

% roman face derivative
\providecommand{\dd}[2]{\ensuremath{\frac{\mathrm{d} #1}{\mathrm{d} #2}}}
\providecommand{\ddinline}[2]{\ensuremath{\mathrm{d} #1/\mathrm{d} #2}}

\ifthenelse{\boolean{cms@italic}}{\newcommand{\cmsSymbolFace}{\relax}}{\newcommand{\cmsSymbolFace}{\mathrm}}

% Particle names which track the italic/non-italic face convention
\providecommand{\zp}{\ensuremath{\cmsSymbolFace{Z}^\prime}\xspace}
\providecommand{\JPsi}{\ensuremath{\cmsSymbolFace{J}\hspace{-.08em}/\hspace{-.14em}\psi}\xspace}
\providecommand{\Z}{\ensuremath{\cmsSymbolFace{Z}}\xspace}
\providecommand{\ttbar}{\ensuremath{\cmsSymbolFace{t}\overline{\cmsSymbolFace{t}}}\xspace}

% Extensions for missing names in PENNAMES
\newcommand{\cPgn}{\ensuremath{\nu}}
\newcommand{\cPJgy}{\JPsi}
\newcommand{\cPZ}{\Z}
\newcommand{\cPZpr}{\zp}

% SM (still to be classified)

\providecommand{\AFB}{\ensuremath{A_\text{FB}}\xspace}
\providecommand{\wangle}{\ensuremath{\sin^{2}\theta_{\text{eff}}^\text{lept}(M^2_\Z)}\xspace}
\providecommand{\stat}{\ensuremath{\,\text{(stat.)}}\xspace}
\providecommand{\syst}{\ensuremath{\,\text{(syst.)}}\xspace}
\providecommand{\kt}{\ensuremath{k_{\mathrm{T}}}\xspace}

\providecommand{\BC}{\ensuremath{\mathrm{B_{c}}}\xspace}
\providecommand{\bbarc}{\ensuremath{\mathrm{\overline{b}c}}\xspace}
\providecommand{\bbbar}{\ensuremath{\mathrm{b\overline{b}}}\xspace}
\providecommand{\ccbar}{\ensuremath{\mathrm{c\overline{c}}}\xspace}
\providecommand{\bspsiphi}{\ensuremath{\mathrm{B_s} \to \JPsi\, \phi}\xspace}
\providecommand{\EE}{\ensuremath{\mathrm{e^+e^-}}\xspace}
\providecommand{\MM}{\ensuremath{\mu^+\mu^-}\xspace}
\providecommand{\TT}{\ensuremath{\tau^+\tau^-}\xspace}

%%%  E-gamma definitions
\providecommand{\HGG}{\ensuremath{\mathrm{H}\to\gamma\gamma}}
\providecommand{\GAMJET}{\ensuremath{\gamma + \text{jet}}}
\providecommand{\PPTOJETS}{\ensuremath{\mathrm{pp}\to\text{jets}}}
\providecommand{\PPTOGG}{\ensuremath{\mathrm{pp}\to\gamma\gamma}}
\providecommand{\PPTOGAMJET}{\ensuremath{\mathrm{pp}\to\gamma + \mathrm{jet}}}
\providecommand{\MH}{\ensuremath{M_{\mathrm{H}}}}
\providecommand{\RNINE}{\ensuremath{R_\mathrm{9}}}
\providecommand{\DR}{\ensuremath{\Delta R}}

%%%%%%
% From Albert
%

\providecommand{\ga}{\ensuremath{\gtrsim}}
\providecommand{\la}{\ensuremath{\lesssim}}
\providecommand{\swsq}{\ensuremath{\sin^2\theta_\cmsSymbolFace{W}}\xspace}
\providecommand{\cwsq}{\ensuremath{\cos^2\theta_\cmsSymbolFace{W}}\xspace}
\providecommand{\tanb}{\ensuremath{\tan\beta}\xspace}
\providecommand{\tanbsq}{\ensuremath{\tan^{2}\beta}\xspace}
\providecommand{\sidb}{\ensuremath{\sin 2\beta}\xspace}
\providecommand{\alpS}{\ensuremath{\alpha_S}\xspace}
\providecommand{\alpt}{\ensuremath{\tilde{\alpha}}\xspace}

\providecommand{\QL}{\ensuremath{\cmsSymbolFace{Q}_\cmsSymbolFace{L}}\xspace}
\providecommand{\sQ}{\ensuremath{\tilde{\cmsSymbolFace{Q}}}\xspace}
\providecommand{\sQL}{\ensuremath{\tilde{\cmsSymbolFace{Q}}_\cmsSymbolFace{L}}\xspace}
\providecommand{\ULC}{\ensuremath{\cmsSymbolFace{U}_\cmsSymbolFace{L}^\cmsSymbolFace{C}}\xspace}
\providecommand{\sUC}{\ensuremath{\tilde{\cmsSymbolFace{U}}^\cmsSymbolFace{C}}\xspace}
\providecommand{\sULC}{\ensuremath{\tilde{\cmsSymbolFace{U}}_\cmsSymbolFace{L}^\cmsSymbolFace{C}}\xspace}
\providecommand{\DLC}{\ensuremath{\cmsSymbolFace{D}_\cmsSymbolFace{L}^\cmsSymbolFace{C}}\xspace}
\providecommand{\sDC}{\ensuremath{\tilde{\cmsSymbolFace{D}}^\cmsSymbolFace{C}}\xspace}
\providecommand{\sDLC}{\ensuremath{\tilde{\cmsSymbolFace{D}}_\cmsSymbolFace{L}^\cmsSymbolFace{C}}\xspace}
\providecommand{\LL}{\ensuremath{\cmsSymbolFace{L}_\cmsSymbolFace{L}}\xspace}
\providecommand{\sL}{\ensuremath{\tilde{\cmsSymbolFace{L}}}\xspace}
\providecommand{\sLL}{\ensuremath{\tilde{\cmsSymbolFace{L}}_\cmsSymbolFace{L}}\xspace}
\providecommand{\ELC}{\ensuremath{\cmsSymbolFace{E}_\cmsSymbolFace{L}^\cmsSymbolFace{C}}\xspace}
\providecommand{\sEC}{\ensuremath{\tilde{\cmsSymbolFace{E}}^\cmsSymbolFace{C}}\xspace}
\providecommand{\sELC}{\ensuremath{\tilde{\cmsSymbolFace{E}}_\cmsSymbolFace{L}^\cmsSymbolFace{C}}\xspace}
\providecommand{\sEL}{\ensuremath{\tilde{\cmsSymbolFace{E}}_\cmsSymbolFace{L}}\xspace}
\providecommand{\sER}{\ensuremath{\tilde{\cmsSymbolFace{E}}_\cmsSymbolFace{R}}\xspace}
\providecommand{\sFer}{\ensuremath{\tilde{\cmsSymbolFace{f}}}\xspace}
\providecommand{\sQua}{\ensuremath{\tilde{\cmsSymbolFace{q}}}\xspace}
\providecommand{\sUp}{\ensuremath{\tilde{\cmsSymbolFace{u}}}\xspace}
\providecommand{\suL}{\ensuremath{\tilde{\cmsSymbolFace{u}}_\cmsSymbolFace{L}}\xspace}
\providecommand{\suR}{\ensuremath{\tilde{\cmsSymbolFace{u}}_\cmsSymbolFace{R}}\xspace}
\providecommand{\sDw}{\ensuremath{\tilde{\cmsSymbolFace{d}}}\xspace}
\providecommand{\sdL}{\ensuremath{\tilde{\cmsSymbolFace{d}}_\cmsSymbolFace{L}}\xspace}
\providecommand{\sdR}{\ensuremath{\tilde{\cmsSymbolFace{d}}_\cmsSymbolFace{R}}\xspace}
\providecommand{\sTop}{\ensuremath{\tilde{\cmsSymbolFace{t}}}\xspace}
\providecommand{\stL}{\ensuremath{\tilde{\cmsSymbolFace{t}}_\cmsSymbolFace{L}}\xspace}
\providecommand{\stR}{\ensuremath{\tilde{\cmsSymbolFace{t}}_\cmsSymbolFace{R}}\xspace}
\providecommand{\stone}{\ensuremath{\tilde{\cmsSymbolFace{t}}_1}\xspace}
\providecommand{\sttwo}{\ensuremath{\tilde{\cmsSymbolFace{t}}_2}\xspace}
\providecommand{\sBot}{\ensuremath{\tilde{\cmsSymbolFace{b}}}\xspace}
\providecommand{\sbL}{\ensuremath{\tilde{\cmsSymbolFace{b}}_\cmsSymbolFace{L}}\xspace}
\providecommand{\sbR}{\ensuremath{\tilde{\cmsSymbolFace{b}}_\cmsSymbolFace{R}}\xspace}
\providecommand{\sbone}{\ensuremath{\tilde{\cmsSymbolFace{b}}_1}\xspace}
\providecommand{\sbtwo}{\ensuremath{\tilde{\cmsSymbolFace{b}}_2}\xspace}
\providecommand{\sLep}{\ensuremath{\tilde{\cmsSymbolFace{l}}}\xspace}
\providecommand{\sLepC}{\ensuremath{\tilde{\cmsSymbolFace{l}}^\cmsSymbolFace{C}}\xspace}
\providecommand{\sEl}{\ensuremath{\tilde{\cmsSymbolFace{e}}}\xspace}
\providecommand{\sElC}{\ensuremath{\tilde{\cmsSymbolFace{e}}^\cmsSymbolFace{C}}\xspace}
\providecommand{\seL}{\ensuremath{\tilde{\cmsSymbolFace{e}}_\cmsSymbolFace{L}}\xspace}
\providecommand{\seR}{\ensuremath{\tilde{\cmsSymbolFace{e}}_\cmsSymbolFace{R}}\xspace}
\providecommand{\snL}{\ensuremath{\tilde{\nu}_L}\xspace}
\providecommand{\sMu}{\ensuremath{\tilde{\mu}}\xspace}
\providecommand{\sNu}{\ensuremath{\tilde{\nu}}\xspace}
\providecommand{\sTau}{\ensuremath{\tilde{\tau}}\xspace}
\providecommand{\Glu}{\ensuremath{\cmsSymbolFace{g}}\xspace}
\providecommand{\sGlu}{\ensuremath{\tilde{\cmsSymbolFace{g}}}\xspace}
\providecommand{\Wpm}{\ensuremath{\cmsSymbolFace{W}^{\pm}}\xspace}
\providecommand{\sWpm}{\ensuremath{\tilde{\cmsSymbolFace{W}}^{\pm}}\xspace}
\providecommand{\Wz}{\ensuremath{\cmsSymbolFace{W}^{0}}\xspace}
\providecommand{\sWz}{\ensuremath{\tilde{\cmsSymbolFace{W}}^{0}}\xspace}
\providecommand{\sWino}{\ensuremath{\tilde{\cmsSymbolFace{W}}}\xspace}
\providecommand{\Bz}{\ensuremath{\cmsSymbolFace{B}^{0}}\xspace}
\providecommand{\sBz}{\ensuremath{\tilde{\cmsSymbolFace{B}}^{0}}\xspace}
\providecommand{\sBino}{\ensuremath{\tilde{\cmsSymbolFace{B}}}\xspace}
\providecommand{\Zz}{\ensuremath{\cmsSymbolFace{Z}^{0}}\xspace}
\providecommand{\sZino}{\ensuremath{\tilde{\cmsSymbolFace{Z}}^{0}}\xspace}
\providecommand{\sGam}{\ensuremath{\tilde{\gamma}}\xspace}
\providecommand{\chiz}{\ensuremath{\tilde{\chi}^{0}}\xspace}
\providecommand{\chip}{\ensuremath{\tilde{\chi}^{+}}\xspace}
\providecommand{\chim}{\ensuremath{\tilde{\chi}^{-}}\xspace}
\providecommand{\chipm}{\ensuremath{\tilde{\chi}^{\pm}}\xspace}
\providecommand{\Hone}{\ensuremath{\cmsSymbolFace{H}_\cmsSymbolFace{d}}\xspace}
\providecommand{\sHone}{\ensuremath{\tilde{\cmsSymbolFace{H}}_\cmsSymbolFace{d}}\xspace}
\providecommand{\Htwo}{\ensuremath{\cmsSymbolFace{H}_\cmsSymbolFace{u}}\xspace}
\providecommand{\sHtwo}{\ensuremath{\tilde{\cmsSymbolFace{H}}_\cmsSymbolFace{u}}\xspace}
\providecommand{\sHig}{\ensuremath{\tilde{\cmsSymbolFace{H}}}\xspace}
\providecommand{\sHa}{\ensuremath{\tilde{\cmsSymbolFace{H}}_\cmsSymbolFace{a}}\xspace}
\providecommand{\sHb}{\ensuremath{\tilde{\cmsSymbolFace{H}}_\cmsSymbolFace{b}}\xspace}
\providecommand{\sHpm}{\ensuremath{\tilde{\cmsSymbolFace{H}}^{\pm}}\xspace}
\providecommand{\hz}{\ensuremath{\cmsSymbolFace{h}^{0}}\xspace}
\providecommand{\Hz}{\ensuremath{\cmsSymbolFace{H}^{0}}\xspace}
\providecommand{\Az}{\ensuremath{\cmsSymbolFace{A}^{0}}\xspace}
\providecommand{\Hpm}{\ensuremath{\cmsSymbolFace{H}^{\pm}}\xspace}
\providecommand{\sGra}{\ensuremath{\tilde{\cmsSymbolFace{G}}}\xspace}
\providecommand{\mtil}{\ensuremath{\tilde{m}}\xspace}
\providecommand{\rpv}{\ensuremath{\rlap{\kern.2em/}R}\xspace}
\providecommand{\LLE}{\ensuremath{LL\bar{E}}\xspace}
\providecommand{\LQD}{\ensuremath{LQ\bar{D}}\xspace}
\providecommand{\UDD}{\ensuremath{\overline{UDD}}\xspace}
\providecommand{\Lam}{\ensuremath{\lambda}\xspace}
\providecommand{\Lamp}{\ensuremath{\lambda'}\xspace}
\providecommand{\Lampp}{\ensuremath{\lambda''}\xspace}
\providecommand{\spinbd}[2]{\ensuremath{\bar{#1}_{\dot{#2}}}\xspace}

\providecommand{\MD}{\ensuremath{{M_\mathrm{D}}}\xspace}% ED mass
\providecommand{\Mpl}{\ensuremath{{M_\mathrm{Pl}}}\xspace}% Planck mass
\providecommand{\Rinv} {\ensuremath{{R}^{-1}}\xspace} 

%% file: QCD-10-023-authorlist.tex
\textbf{Yerevan Physics Institute,  Yerevan,  Armenia}\\*[0pt]
V.~Khachatryan, A.M.~Sirunyan, A.~Tumasyan
\vskip\cmsinstskip
\textbf{Institut f\"{u}r Hochenergiephysik der OeAW,  Wien,  Austria}\\*[0pt]
W.~Adam, T.~Bergauer, M.~Dragicevic, J.~Er\"{o}, C.~Fabjan, M.~Friedl, R.~Fr\"{u}hwirth, V.M.~Ghete, J.~Hammer\cmsAuthorMark{1}, S.~H\"{a}nsel, C.~Hartl, M.~Hoch, N.~H\"{o}rmann, J.~Hrubec, M.~Jeitler, G.~Kasieczka, W.~Kiesenhofer, M.~Krammer, D.~Liko, I.~Mikulec, M.~Pernicka, H.~Rohringer, R.~Sch\"{o}fbeck, J.~Strauss, A.~Taurok, F.~Teischinger, P.~Wagner, W.~Waltenberger, G.~Walzel, E.~Widl, C.-E.~Wulz
% ...
\vskip\cmsinstskip
% ...
1:~~Also at CERN, European Organization for Nuclear Research, Geneva, Switzerland\\
% ...
